\documentclass[12pt]{iopart}

\usepackage{setstack}
\usepackage{lscape}
\usepackage{epsfig}
\usepackage{rotating}
\usepackage{cite}
\begin{document}

\title[Continuum hard rod gases vs.\ discrete exclusion processes]{Hard rod gas with long-range interactions: Exact predictions for hydrodynamic properties of continuum systems from discrete models}

\author{G. Sch{\"o}nherr\footnote[1]{To
whom correspondence should be addressed (g.schoenherr@fz-juelich.de)}}

\address{Institut f{\"u}r Festk{\"o}rperforschung, Forschungszentrum J{\"u}lich, 52425 J{\"u}lich, Germany}

\begin{abstract}
One-dimensional hard rod gases are explicitly constructed as the limits of discrete systems: exclusion processes involving particles of arbitrary length. Those continuum many-body systems in general do not exhibit the same hydrodynamic properties as the underlying discrete models. Considering as examples a hard rod gas with additional long-range interaction and the generalized asymmetric exclusion process for extended particles ($\ell$-ASEP), it is shown how a correspondence between continuous and discrete systems must be established instead. This opens up a new possibility to exactly predict the hydrodynamic behaviour of this continuum system under Eulerian scaling by solving its discrete counterpart with analytical or numerical tools. As an illustration, simulations of the totally asymmetric exclusion process ($\ell$-TASEP) are compared to analytical solutions of the model and applied to the corresponding hard rod gas. The case of short-range interaction is treated separately.
\end{abstract}


\maketitle

\section{Introduction and outline}
\label{Intro}
Modeling stochastic many-body systems is fundamental to the investigation of driven diffusive systems \cite{lig,kip,spo,priv,sch2}. So far, only few models have been solved analytically. With growing computer power, numerical tools like Monte Carlo (MC) simulations of discrete lattice models steadily gain in importance. However, for many physical processes, the restriction to discrete space is artificial, and the question arises, to which extent discretized models lead to a valid description of experimental settings taking place in continuous space. In the present paper, one-dimensional continuum models are investigated as to their explicit construction from discrete systems. The continuum gases treated in the following are one-dimensional hard rod gases exposed to an external driving field. Two types of hard rod gases are presented: The first consists of hard rods which do not show any interaction besides exclusion. This continuum gas is obtained as the limit of the discrete generalized asymmetric exclusion process for extended particles ($\ell$-ASEP) \cite{schoen}. Basic properties of the discrete model and its hydrodynamic (HD) limit are summarized in section \ref{ASEP} before carrying out the construction of its continuum limit explicitly in section \ref{rescaling}. Sections \ref{hdl}-\ref{TASEPsolutions} are devoted to the investigation of the HD properties of the hard rod gas which turn out very differently from those of its discrete origin, the $\ell$-ASEP. The general proceedings of sections \ref{ASEP} and \ref{hardrodgas} are repeated for the construction and investigation of a second type of hard rod gas in section \ref{longrange}: hard rods with an additional long-range interaction. In subsections \ref{lasepzrp}-\ref{cdcontinuum} such a continuum gas with long-range interaction is realized as the limit of an exclusion process with headway-dependent hopping rates. This leads to remarkable results about the nature of discrete and continuum systems. Analysis shows that the hydrodynamic properties of this continuum gas are exactly the same as for the discrete $\ell$-ASEP with constant stochastic rates. In section \ref{interactiontype}, the common macroscopic features of the $\ell$-ASEP and the long-range interacting continuum hard rod gas concerned are examined for their physical, microscopic causes. The usefulness of the correspondence established between those two systems is illustrated in section \ref{simulations} where results from lattice MC simulations are applied to the continuum hard rod gas.

\section{A discrete model: the generalized asymmetric exclusion process ($\ell$-ASEP)}
\label{ASEP}

\begin{figure}[h]
\begin{center}
\epsfig{file=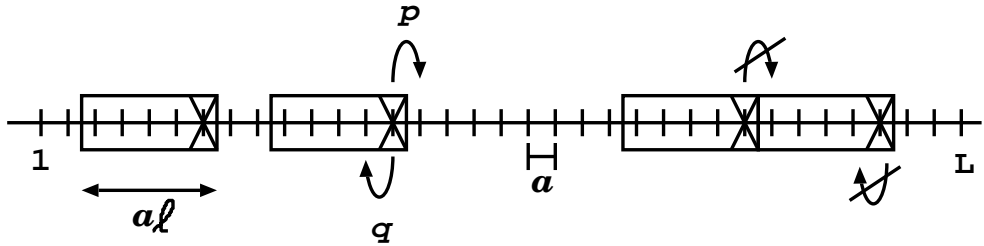, height=\linewidth, angle=270}
\caption{\label{lASEP}The $\ell$-ASEP: extended particles cover $\ell$ lattice sites each and move via next-neighbour hopping with stochastic hopping rates $p$ and $q$.}
\end{center}
\label{rodspoints}
\end{figure}

The $\ell$-ASEP is a stochastic lattice model for a system of extended particles which interact via exclusion and which are exposed to a driving field. The microscopic dynamics of the model define hopping processes of the particles taking place between neighbouring lattice sites with certain stochastic rates. The parameter $\ell$ fixes the length of the particles on lattice scale. For the case $\ell=1$, the $\ell$-ASEP reduces to the well-known asymmetric exclusion process (ASEP) \cite{sch2}. Figure \ref{lASEP} illustrates the hopping dynamics for a lattice of $L(=35)$ sites, $k=1 \ldots L$, which contains $N(=4)$ particles, each covering $\ell(=5)$ adjacent lattice sites. The position of each particle is by convention taken as the last site that the particle occupies towards the right handside of the chain, and is marked with a cross. A particle may change its position from a site $k$ to $k+1$ with a rate $p$ provided that site $k+1$ is empty. Vice versa, it may hop from site $k$ to $k-1$ with a stochastic rate $q$ provided that site $k-\ell$ is not occupied by any particle.
In the thermodynamic limit the number $N$ of particles and the number $L$ of lattice sites tend to infinity while the particle density remains constant. On a coarse-grained scale the lattice constant $a$ is a comparably small quantity. In a continuum limit $a \rightarrow 0$, the discrete site label $k$ may be replaced by a continuous coordinate $x$. The evolution of the local particle density $\rho(x,t)$ is determined by a continuity equation of the form \cite{schoen}\\
\begin{equation}
\partial_t \rho = - a(p-q)\partial_x[v\rho] - a^2\frac{p+q}{2}\partial_{xx}v + O(a^3)\,.
\label{HD}
\end{equation}
where $v$ is a normalized average particle velocity \cite{schoen,fer}
\begin{equation}
v=\frac{1-\ell \rho}{1-(\ell-1)\rho}\,.
\end{equation}
It must be stressed that this equation is an approximation for a system with discrete dynamics and that it does not describe the motion of hard rods with dynamics defined in continuous space.
The $\ell$-ASEP was originally introduced as a lattice model for protein synthesis in 1968 \cite{mac1,mac2} and has been further investigated in this context recently \cite{lak,shaw1,shaw2,shaw3}. In those works, the extended particles, hopping stochastically along a chain, represent the motion of ribosomes along the codons of a m-RNA template. For this purpose the description of the system by a discrete model is appropriate because the codons form a real biological discrete lattice. Seeking to apply the $\ell$-ASEP to model a greater class of physical experiments which take place in continuous space (e.g. the study of colloidal suspensions), a continuum hard rod gas is constructed as a limit of the $\ell$-ASEP.\\

\section{Continuum hard rod gas}
\label{hardrodgas}

\subsection{Construction of a hard rod gas: rescaling the particle length}
\label{rescaling}

The derivation of the hydrodynamic equation (\ref{HD}) of the $\ell$-ASEP involves a continuum limit, in which the lattice constant $a$ approaches zero. This equation in general does not describe the dynamics of a gas of hard rods in continuous space. The physical extension of the $\ell$-ASEP particles $\ell a$ vanishes on a macroscopic scale. The particles appear like point particles from a coarse-grained point of view.\\
In order to observe rods of non-zero extension in the continuum limit of space, their length on lattice scale must be rescaled during the limiting procedure as:
\begin{equation}
\ell = \frac {\lambda}{a} \, .
\end{equation}
$\lambda$ is held constant for $a \rightarrow 0$ and indicates the true length of the rods in physical units. The continuum limit is performed for a fixed number of rods of length $\lambda$ on a chain of length $\Lambda$ ($L \rightarrow \infty$ sites). As the number $\ell$ of sites which are occupied by one particle diverges for $a \rightarrow 0$, the particle density $\rho$ on lattice scale tends to zero, while the coverage density $\rho^{c}=\ell \rho$ and the volume particle density $\tilde \rho = \frac{\rho^{c}}{\lambda}$ remain constant. This \emph{hard rod limit} of the $\ell$-ASEP is illustrated in figure \ref{continuumlimit}(b),{\bf{A}}$\rightarrow${\bf{B}}.

\subsection{Hydrodynamic limit}
\label{hdl}
The hydrodynamic limit of the continuum hard rod system is realized in a second step, considering a big system at constant density ($\Lambda \rightarrow \infty, \rightarrow \infty, \frac{N}{\Lambda}\equiv const.$). During a second coarse-graining process, the length $\lambda$ of the rods becomes a comparably small quantity (compare figure \ref{continuumlimit} (b),{\bf{B}}$\rightarrow${\bf{C}}). The hydrodynamic equation of the continuum hard rod gas may be deduced as the hard rod limit of the HD equation (\ref{HD}) of the discrete system (in figure \ref{continuumlimit}: (a),{\bf{B}}$\rightarrow$(b),{\bf{C}}).
As the discrete particle density $\rho$ becomes zero in the hard rod limit, the coverage density $\rho^c$ will be considered instead.
In terms of $\rho^c$, and employing the abbreviations $B=p-q$ and $S=\frac{p+q}{2}$, equation (\ref{HD}) is transformed into:

\begin{eqnarray}
\partial_t \rho^c = -aB \partial_x \left[ \frac{\rho^c (1- \rho^c )}{1-\rho^c + \frac{\rho^c}{\ell}} \right] + a^2 S \partial_{xx} \left[ \frac{\rho^c}{1-\rho^c + \frac{\rho^c}{\ell}} \right] + O(a^3) \nonumber \\
                  = -aB\partial_x \rho^c + a^2 \left[ B \frac{(\rho^c)^2}{\lambda (1-\rho^c)} + S \partial_{xx}  \frac{\rho^c}{1-\rho^c} \right] + O(a^3) \qquad {\rm (if \rho^c \ne 1)} \nonumber \\
                  = -\partial_x \left[ j(\rho^c) \right].
\label{particlehd}
\end{eqnarray}

\begin{figure}[b]
\centering\epsfig{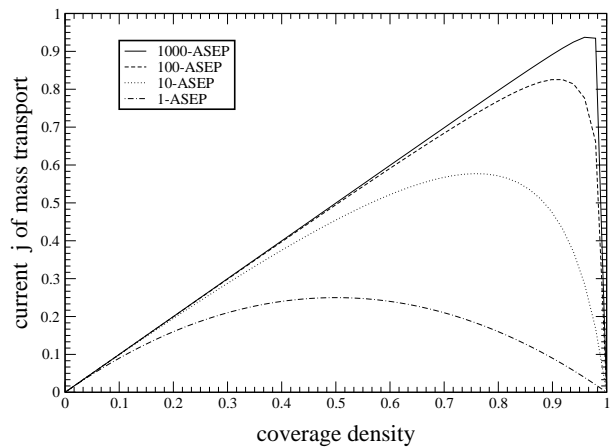}
\caption{\label{plotj}The $\ell$-ASEP mass current as a function of the coverage density $\rho^c$:
Asymptotically ($\ell \rightarrow \infty$), the current-density relation becomes linear.}
\end{figure}

The scaling $\ell=\frac{\lambda}{a}$ and the limit of small $a$ have been used to expand the terms. Furthermore (\ref{particlehd}) is only valid for the case of non-maximal density, which has been assumed in the first step in order to be able to perform the approximation without yielding singularities. The case of maximal density will be discussed in section \ref{TASEPsolutions}.\\
Under Eulerian scaling $[\partial_t \equiv a \partial_x]$, all terms in the current of orders higher than $a$ are negligible. The evolution of the coverage density is determined by
\begin{equation}
\partial_t \rho^c (x,t)+ B \partial_x \rho^c (x,t) = 0 \, , \qquad \rho^c \ne 1 \,.
\label{hdrods}
\end{equation}
The asymptotic approach of the mass current as a function of the coverage density from a convex to a linear function is demonstrated in figure \ref{plotj}. The linearity of the current in the hard rod limit implies that all rods are moving with constant velocity
\begin{equation}
v=\frac {\partial j}{\partial \tilde \rho} = B = (p-q)
\end{equation}
into the direction of the drive. In contrast to the discrete $\ell$-ASEP, interaction effects disappear completely.

\subsection{Solutions of the $\ell$-TASEP in its hydrodynamic and hard rod limit}
\label{TASEPsolutions}

Equation (\ref{hdrods}) excludes the case of particle distributions, where the density assumes its maximum at some point. In this section, explicit solutions of the Eulerian current of the $\ell$-ASEP will be constructed. Their asymptotic behaviour in the continuum hard rod limit is discussed in particular for the case of partly maximal initial density profiles. For the sake of simplicity of notation the $\ell$-ASEP shall be limited to its totally asymmetric case, the so-called $\ell$-TASEP, whose hopping rates equal $p=1$ and $q=0$.\\
\\
On Euler scale, only the lowest order terms of (\ref{HD}) are taken into account for the $\ell$-TASEP current
\begin{equation}
j(\rho)= \frac{\rho(1-\ell \rho)}{1-(\ell -1)\rho}
\end{equation}
which obeys the continuity equation
\begin{equation}
\partial_t \rho(x,t) + \partial_x j(\rho) = 0 \,.
\label{pde1}
\end{equation}
This partial differential equation (PDE) governs the evolution of the average particle fraction $\rho$ per lattice site. In order to obtain the picture of macroscopic rods, it is convenient to replace $\rho$ by the volume particle density $\tilde \rho$. Furthermore every term $\ell a$ is replaced by the constant length $\lambda$. Equation (\ref{pde1}) is thus transformed into:

\begin{equation}
 \partial_t \tilde \rho(x,t) + \partial_x j(\tilde \rho) = 0
\end{equation}
where
\begin{equation}
j(\tilde \rho)=  \frac{\tilde \rho(1-\lambda \tilde \rho)}{1- \lambda \tilde \rho + a \tilde \rho}\,.
\end{equation}
In the following, solutions $u(x,t)$ of this PDE of the general form
\begin{eqnarray}
u_t+[f(u)]_x &=& 0  \nonumber \\
f(u)&=& \frac{u(1-\lambda u)}{1-\lambda u + au} \nonumber \\
u(x,0) &=& \phi (x)\,.
\label{pde2}
\end{eqnarray}
are constructed for certain initial profiles $\phi (x)$. In the limit $a \rightarrow 0$ the solutions are solutions of the continuum hard rod gas.\\
\\
First, choose initial conditions of the form:
\begin{equation}
\phi (x) = \left\{ \begin{array}{r@{\quad:\quad}l}
L_0 & x < 0 \\
R_0 & x \ge 0. \end{array} \right.
\label{profil}
\end{equation}
The initial densities $L_0$ and $R_0$ are fractions of the maximal density:
\begin{eqnarray}
L_0 &=& \frac{1}{k_L \lambda } \nonumber \\
R_0 &=& \frac{1}{k_R \lambda } \qquad k_L,k_R \in \Re \,.
\end{eqnarray}
The evolution of initial profiles which have a discontinuity are of special interest because they determine the possibility of the formation of stable shocks in the system. One therefore demands $k_L \ne k_R$. Solutions of problem (\ref{pde2}) with the initial condition (\ref{profil}) are so-called weak solutions which  are piecewise smooth and which obey (\ref{pde2}) only locally.\\
\\
Next, take a look at some general properties of $f(u)$. Taking into account the physical condition $u \le \frac{1}{\lambda}$ for any solution of problem (\ref{pde2}), $f$ proves to be a convex function ($f'(u)<0$).
The convexity of $f$ implies that the unique piecewise smooth solution of (\ref{pde2}),(\ref{profil}) is given by \cite{Levans}:

\begin{equation}
u(x,t) =
\left\{ \begin{array}{r@{\quad:\quad}l}

\left\{ \begin{array}{r@{\quad:\quad}l}
L_0 & x \le  \frac{f(L_0)-f(R_0)}{L_0-R_0}t \\
R_0 & x > \frac{f(L_0)-f(R_0)}{L_0-R_0}t
\end{array} \right. & L_0 < R_0\\

\left\{ \begin{array}{r@{\quad:\quad}l}
L_0 & x \le  f'(L_0)t \\
h(\frac{x}{t}) & f'(L_0)t < x \le f'(R_0)t \\
R_0 & x > f'(R_0)t
\end{array} \right. & L_0 > R_0

\end{array} \right.
\end{equation}
where
\begin{eqnarray}
h = (f')^{-1}\\
\frac{x}{t} =y  \mapsto  u
\end{eqnarray}
is obtained from inverting the first derivative of $f$ and only admitting types of solutions where $f'(u)=y = \frac{x}{t}$ as
\begin{equation}
h(y)= \frac{1}{\lambda -a} - \sqrt{\frac{1}{(\lambda -a)^2} - \frac{y-1}{(\lambda -a)[(\lambda - a)y - \lambda]}} \,.
\label{rarefactioncurve}
\end{equation}
The evolution of the system depends on the initial conditions: An initial discontinuity evolves as a stable shock for $L_0<R_0$, while for $L_0>R_0$, it dissolves in time according to a rarefaction wave solution. From a microscopic viewpoint this means that for the former case fluctuations which arise near the discontinuity are driven towards it according to the Rankine-Hugoniot jump condition for shock stability \cite{Levans} $v_{left}>v_{shock}>v_{right}$ for the local velocities around the discontinuity. In the latter case, fluctuations are driven away from the point of the (initial) discontinuity in the density profile and tend to soften it \cite{sch2}.\\
\\
\textbf{\underline{Limiting case $a \rightarrow 0$}}\\
So far, the solutions of the $\ell$-TASEP current on Euler scale have been considered for finite $a$. In order to apply the solutions to a system of hard rods, the limit $a \rightarrow 0$ is carried out in the following. For volume particle densities smaller than the maximum ($\tilde \rho < \frac{1}{\lambda }$), one expects stability of any initial profile in accordance to equation (\ref{hdrods}). However, it still remains an open question how an initial profile of partly maximal density evolves. The specifications $k_L=1$ and $k_R=1$ are therefore listed separately in the following.\\
\\
\underline{\bf Case 1 ($L_0< R_0):$}
\begin{itemize}
\item For $L_0,R_0 \ne \frac{1}{\lambda } \, (k_L,k_R \ne 1)$ the speed at any point of any  profile is the same. Especially for (\ref{profil}):
\begin{equation}
\lim_{a \to 0} f'(\frac{1}{k\lambda }) \rightarrow 1 (k \in \{k_L,k_R\}) \,.
\end{equation}
Thus the discontinuity is moving with the velocity $v_{shock}=1$.
\item For $L_0<R_0= \frac{1}{\lambda } \, (k_L > k_R =1)$ the slope of the characteristics of the PDE (\ref{pde2}) starting at $x>0$ for $t=0$ approaches 0, which corresponds to an infinitely fast propagation of any fluctuations in the initial high density regime towards the left. As the velocity at the right hand-side of the shock position, $v_{right}$, tends to minus infinity, the Rankine-Hugoniot jump condition is still satisfied. Thus even in the case of maximum density any initial profile is stable. The apparently infinitely fast spreading of the discontinuity has to be interpreted on lattice scale where any hard rod occupies infinitely many sites which have to be covered each time the fluctuation moves.
\end{itemize}

\begin{figure}
\begin{center}
\epsfig{file=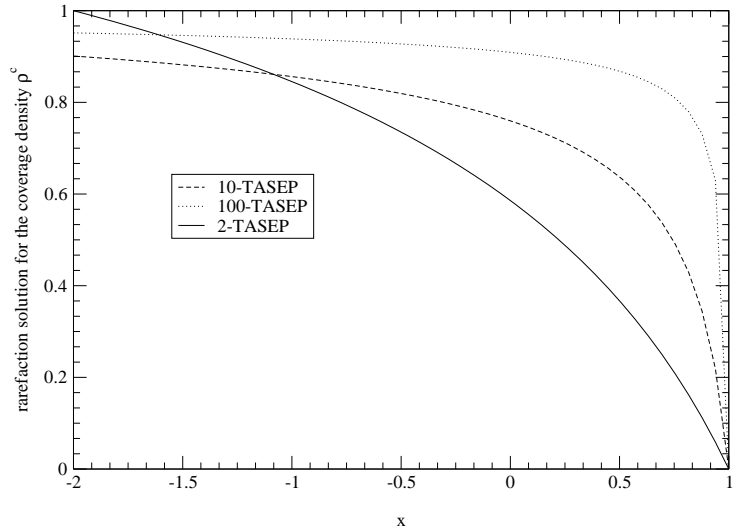, angle=0, height=8cm}
\caption{The rarefaction solutions $\ell \, h_{\ell}(\frac{x}{t})$ of the coverage density of the $\ell$-TASEP for different values of $\ell$ ($\ell=2,10,100$) are plotted at time $t=1$ over a certain x-range close to the zero-density regime ($\rho^c(x,1)=0$ for $x>1$). The initial coverage density was chosen as a step profile ($\rho^c=1$ if $x<0$ and $\rho^c=0$ if $x \ge 0$). The steepness of the curves increases with the length $\ell$ of the particles.}
\label{zrpasep}
\end{center}
\end{figure}

\underline{\bf Case 2 ($L_0>R_0$):}
\begin{itemize}
\item The case $L_0,R_0 \ne \frac{1}{\lambda } \, (k_L,k_R \ne 1)$ is similar to case 1, as the characteristics approach parallelism again, and the initial profile moves with constant velocity. The construction of $h(y)$ is not necessary, because there is no space-time area which is not covered by the characteristics.
\item For $R_0<L_0= \frac{1}{\lambda } \, (k_R > k_L =1)$, the asymptotic behaviour of the rarefaction solution $\lim_{a \to 0}h(y)$ has to be considered for the space-time wedge $\lim_{a \to 0}f'(\frac{1}{\lambda })=\lim_{a \to 0}-\frac{\lambda }{a}=-\infty <y \le 1=\lim_{a \to 0} f'(\frac{1}{k_R \lambda })$.
Expanding $h(y,a)$ in orders of $a$ yields:
\begin{equation}
h(y,a)= \frac{1}{\lambda } - \frac{\sqrt{-\frac{1}{\lambda (y-1)}}}{\lambda}\sqrt{a} + \frac{a}{\lambda^2}+ O(a^{\frac{3}{2}})\,.
\end{equation}
Thus $h$ approaches a constant function in the limit $a \rightarrow 0$:
\begin{equation}
\lim_{a \to 0} h(y,a) = \frac{1}{\lambda }\,.
\end{equation}
This means that the rarefaction wave solution approaches a solution containing a sharp jump in the limit, and that the discontinuity remains stable in this case as well (compare fig. \ref{zrpasep}).
\end{itemize}
Summing this up, the predicted result, i.e.\ a linear current in the hard rod limit, has been confirmed for the case of profiles of less than maximal density and its validity could also be enlarged to the case of piecewise  maximal initial density profiles.\\

\section{Hard rods with long-range interaction}
\label{longrange}
Long-range interaction is introduced in the following as a new feature in the framework of the discrete $\ell$-ASEP. It is shown, how a generalized form of the $\ell$-ASEP with headway-dependent hopping rates arises naturally from a mapping between exclusion processes and a certain class of zero range processes (ZRP) \cite{spit,evan}. A hard rod gas with long-range interaction is constructed as the continuum hard rod limit of those processes.

\subsection{The $\ell$-ASEP with headway-dependent rates and zero range processes}
\label{lasepzrp}
\begin{figure}
\begin{center}
\epsfig{file=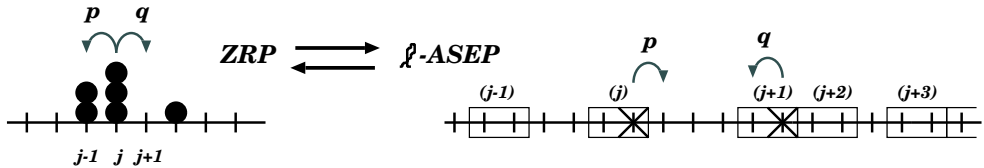, height=\linewidth, angle=270}
\caption{\label{mapping}Concept of the mapping between ZRP and $\ell$-ASEP: ZRP lattice sites are turned into particles, ZRP particles are replaced by holes; the stochastic hopping rates $p$ and $q$ are interchanged.}
\end{center}
\end{figure}

Like the $\ell$-ASEP, a zero range process describes the stochastic hopping of particles on a lattice. However there are important differences in its microscopic definition: Firstly, the particles have zero interaction range, i.e. there is no restriction of the number of particles which may be located at the same lattice site. Secondly, the hopping events in general can take place between any two sites $i$ and $i'$ of the ZRP lattice and depend on the occupation numbers of those sites. In the following, a ZRP will be considered where the particles move along the lattice via next-neighbour hopping with rates $p\,\omega_n$ to the left and $q\, \omega_n$ to the right, where $\omega_n$ defines the occupation number dependence of the corresponding hopping rate and $p-q$ determines the overall hopping asymmetry. For the case of constant rates $\omega_n \equiv \omega$, it has been described in \cite{schoen} how to construct a one-to-one mapping between ZRP and $\ell$-ASEP explicitly. The basic idea is to replace ZRP-sites by $\ell$-ASEP particles and ZRP-particles by $\ell$-ASEP holes (compare figure \ref{mapping}). The densities $\rho_k$ of the extended particles and $c_i$ of the ZRP particles are related as follows:
\begin{equation}
\rho_k= \frac{1}{c_i+\ell}
\label{dtrafo}
\end{equation}
where the discrete coordinates of the two systems transform like
\begin{equation}
k = \left( \sum_{i'=0}^{i-1} c_i'(t) \right)+ i \ell \,.
\label{trafo}
\end{equation}
Application of transformations (\ref{dtrafo}) and (\ref{trafo}) to ZRP with occupation number dependent rates $\omega_n$ yields a system of extended particles with exclusion interaction and headway-dependent hopping rates.

\subsection{The zero range process: stationary state and hydrodynamic equation}
\label{zrpproperties}
The basic properties of the ZRP shall be shortly summarized in this section.
The stationary state of a zero range process on a lattice of $N$ sites $j=1\cdots N$ with rates $\omega_{\eta(j)}$ which depend only on the occupation $\eta(j)$ of the site of departure, is known to factorize into a product measure \cite{spit}:
\begin{equation}
P^*(\eta(1),\eta(2),\cdots,\eta(N))=f(\eta(1))f(\eta(2))\cdots f(\eta(N))
\end{equation}
where
\begin{equation}
f(n)= \frac{p_n z^n}{\sum_{k=0}^{\infty}p_n z^n},\quad n=0,1,2,\cdots
\end{equation}
with
\begin{eqnarray}
   z= \sum_{n=1}^{\infty}f_n\omega_n\\
 p_n= \prod_{m=1}^n \frac{1}{\omega_m}, \qquad p_0=1
\end{eqnarray}
denote the single site probabilities for occupancy  by $n$ particles and are represented by power series in the fugacity $z$.\\
\\
Introducing a non-equilibrium analogue of the partition function
\begin{equation}
Z=\sum_{n=0}^{\infty} p_n z^n
\label{partition}
\end{equation}
the ZRP density $c$ can be deduced from the fugacity $z$ as
\begin{equation}
c=z\frac{d}{dz}ln Z
\label{cz}
\end{equation}
In the HD limit, the fugacity and the density become smooth functions of a continuous space coordinate $y$.
The hydrodynamic equation for the density evolution of the ZRP, derived from the master equation of the stochastic process, takes the form \cite{schoen}:
\begin{equation}
\partial_t c(y,t)=aB\partial_y z(y,t) + a^2S\partial_{yy} z(y,t) + O(a^3) \,.
\label{HDZRP}
\end{equation}
Calculating (\ref{cz}) for a certain choice of rates $\omega_n$ yields a hydrodynamic equation in terms of the ZRP density $c(y,t)$.

\subsection{Construction of a continuum hard rod gas with long-range interaction}
\label{cdcontinuum}
As outlined above, a class of ZRP with occupation-number dependent rates through a one-to-one mapping induces a class of exclusion processes where the hopping rates of the involved particles are a function of their headway. This mapping is carried out explicitly in the appendix. Its implications for the investigation of continuum systems become evident in the present section where a hard rod gas with long-range interaction is yielded as the continuum hard rod analogue of a ZRP with occupation-number dependent rates.\\
\\
Applying substitutions of densities and space coordinates arising from (\ref{dtrafo}) and (\ref{trafo}) to equation (\ref{HDZRP}) yields the hydrodynamic equation for extended particles with exclusion interaction and headway dependent hopping rates of the general form (compare appendix, equation (\ref{vZRPgeneral})):
\begin{equation}
\partial_t \rho + aB\partial_x[\rho v(\rho)] + a^2S\partial_{xx}v(\rho)=0
\label{genHDE}
\end{equation}
where the velocity term $v(\rho(x))=z(c(\rho(x)))$ is identical to the transformed expression of the ZRP fugacity. The hard rod limit of (\ref{genHDE})in terms of the volume particle density $\tilde \rho =\frac{\rho}{a}$ under Eulerian scaling is calculated as (compare section \ref{hdl}):
\begin{eqnarray}
\partial_t \tilde \rho &=&  -B\partial_x [\tilde \rho z(c(a\tilde \rho))]
\label{hdfugacity}
\end{eqnarray}
where
\begin{equation*}
c(\rho)=c(a\tilde \rho)=\frac{1}{a}(\frac{1}{\tilde \rho}-\lambda) \, .
\end{equation*}
The form of the current-density relation for the hard rods is therefore completely determined by the fugacity-density relation (\ref{cz}) of the corresponding zero range process, i.e. by the choice of the form of the microscopic ZRP hopping rates $\omega_n$.

\subsection{Current-density relation: determining the stochastic rates}
\begin{figure}
\begin{center}
\epsfig{file=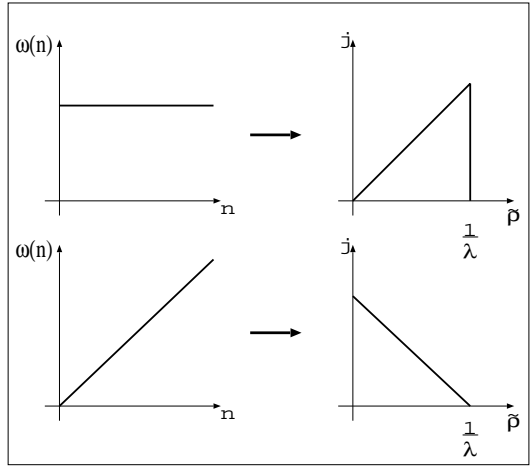, width=7cm, angle=0}
\caption{Qualitative dependence of the current-density relation $j(\tilde \rho)$ of the continuum hard rod gas in its HD limit on the form of the corresponding microscopic ZRP hopping rates $\omega(n)$.}
\end{center}
\label{somerates}
\end{figure}

Having once established equation (\ref{hdfugacity}), the rates of the ZRP which determine the basic relations (\ref{partition}, \ref{cz}) may be manipulated such that the corresponding hard rod gas shows a certain type of coarse-grained current-density relation. A hard rod gas exhibiting hydrodynamic phenomena similar to the ones of the $\ell$-ASEP, for instance stable shocks, would be of much interest. Therefore one seeks to achieve a $\ell$-ASEP type nonlinear current-density relation from equation (\ref{hdfugacity}) as a transformation from a zero range process with appropriately adapted stochastic hopping rates.
For this purpose, the ZRP-rates $\omega_n$ must induce an interaction taking place on \emph{macroscopic} distances between any two rods.
A convex current of $\ell$-ASEP type can only be obtained for rates that increase with the distance of the particles which in turn is proportional to the number of free lattice sites $n$ in front of a particle (i.e. the rates decrease with the particle density).\\
\\
The most simple choice of non-constant rates would be the one where $\omega_n$ increases linearly in $n$ or, if scaling is already taken care of, linearly in $na$. Calculation of the fugacity $z(c(a\tilde \rho))=ac(a\tilde \rho)=\frac{1}{\tilde \rho}-\frac{1}{\lambda}$, however, shows that the resulting current $j= \tilde \rho z$ is also linear (and decreasing) in $\tilde \rho$. Though rates of higher order in $n$, for instance $\omega_n=an^2$, lead to a nonlinear current, such a current is concave and has a pole at $\rho=0$. Figure \ref{somerates} shows how the qualitative properties of the $\lambda$-rod current depend on the choice of the ZRP rates as functions of $n$.\\
\\
The desired current, i.e.\ a current which resembles the $\ell$-ASEP current also in the limit $\ell \rightarrow \infty$, should increase with the density for low system densities and it should decrease with $\rho$ for high densities. Therefore rates are necessary that are nearly constant for big macroscopic distances $r$ and nearly linear in $n$ for small macroscopic distances $r=na$. A possible choice is:
\begin{equation}
\omega_r=\frac{r}{r+1}=\frac{na}{na+1}\,.
\label{rates}
\end{equation}
The derivation of $c(z)$ for such rates (\ref{rates}) is straightforward according to the relations given in section \ref{zrpproperties}. The power series in $z$ constituting the partition function is evaluated as

\begin{eqnarray*}
Z&=&\sum_{n=0}^{\infty}\frac{1}{n!} \left( \prod_{m=1}^{n} \left( \frac{a+1}{a}+m-1 \right) \right) z^n\\
 &=&(1-z)^{-\frac{a+1}{a}}\,.
\end{eqnarray*}
The fugacity $z$ may be given explicitly as a function of $c$. From
\begin{equation*}
c=z \frac{d}{dz}\ln(Z)=\frac{a+1}{a}\frac{z}{1-z}
\end{equation*}
follows:
\begin{equation*}
z(c)=\frac{c}{\frac{a+1}{a}+c}=\frac{ac}{a+1+ac}\,.
\end{equation*}
In the limit $a \rightarrow 0$ and for constant volume particle density $\tilde \rho$ the ZRP density $c$ diverges like $\frac{1}{a}$. The fugacity in this limit approaches:
\begin{eqnarray*}
\lim_{a \to 0}z(c(\tilde \rho))&=&\lim_{a \to 0}\frac{a\frac{1}{a}(\frac{1}{\tilde \rho}-\lambda)}{a+1+a\frac{1}{a}(\frac{1}{\tilde \rho}-\lambda)}\\
                                 &=&\frac{1-\lambda \tilde \rho}{1-(\lambda-1)\tilde \rho}\,.
\end{eqnarray*}
The hydrodynamic equation (\ref{hdfugacity}) applied to interacting hard rods under Eulerian scaling transforms into:
\begin{equation}
\partial_t \tilde \rho = - \partial_x  B \left[ \frac{\tilde \rho (1-\lambda \tilde \rho)}{1-(\lambda - 1)\tilde \rho} \right].
\label{longrangecurrent}
\end{equation}
Remarkably, one recovers exactly the same type of current-density relation as derived for the $\ell$-ASEP before taking the limit $\ell \rightarrow \infty$ of macroscopic rods of rescaled length $\lambda=\ell a$.

\section{Continuum gases and their lattice counterparts}
After a so far mainly formal derivation, a rather phenomenological approach to the comparison of discrete and continuum systems is taken in the following. Results from the preceding sections are reviewed focusing on their physical meaning as well as their practical significance for future calculations of similar systems.

\subsection{Types of interaction: $\ell$-ASEP and hard rod gas}
\label{interactiontype}
The coarse-grained behaviour of the continuum hard rod system with long-range interaction corresponds to the one of a lattice model of extended particles with exclusion interaction. The current-density relations in the continuum limit of both systems are formally equivalent. In the following, it is illustrated why those two systems macroscopically evolve in the same fashion while they are governed by two different kinds of microscopic interactions. The procedure of coarse-graining during the HD limit is reviewed in greater depth.\\
\\
\underline{\bf Long-range interaction}\\
First of all, it shall be demonstrated that a long-range type of interaction is necessary for a hard-rod system with macroscopic interactions.  The limit of vanishing lattice constant $a \rightarrow 0$ is compared for the $\ell$-ASEP and the hard rod gas.
\begin{itemize}
\item When performing the HD limit of the $\ell$-ASEP, the real distance between two particles shrinks but the number of lattice sites between them remains constant. An interaction between two particles takes place, whenever a hopping attempt is rejected because neighbouring sites are occupied. Due to the finite distance, two particles which are separated only by holes in between, in general interact within a finite time.
\item Letting $a$ approach zero for the hard rod gas, both the length of the rods and the inter-particle distances are rescaled like $\frac{1}{a}$. The real distance between two rods remains constant in this case, but the number of vacant lattice sites in between two rods approaches infinity.\\
During the second coarse-graining, the real distance becomes comparably small again, but still contains infinitely many sites of the original lattice. Dynamics which are only defined on the microscopic lattice scale therefore can not possibly lead to any interaction. Two rods would need an infinite amount of time to meet and the sheer exclusion interaction does not play any role for the evolution of the system.\\
If interaction effects are desired, one \emph{must} equip the rods with some kind of long-range interaction which makes them feel the presence of each other even when on lattice scale at an infinite distance. The hopping rates (\ref{rates}) have been scaled to a form such that they are functions of the macroscopic distance $r$ instead of the diverging number of free lattice sites.
\end{itemize}
\underline{\bf Specific form of interaction: ZRP rates $\omega_r = \frac{r}{r+1}$}\\
Having understood the scales of interaction, its specific form shall be investigated. The long-range interaction of the hard rod gas is determined completely by the stochastic hopping rates of the corresponding zero range process and the discrete exclusion process. In the following it is explained why a certain coarse-grained hard rod current, in particular a $\ell$-ASEP type current, must arise as a consequence of those rates.\\
\\
Consider a ZRP with a general form of hopping rates $\omega(r)$ which depend on the inter-particle distance $r$ of the corresponding hard rod system like
\begin{equation}
\omega(r)=\frac{r}{r+x} \, .
\label{generalrates}
\end{equation}
Those rates result in a current-density relation of the form:
\begin{equation}
j=\frac{\tilde \rho (1-\lambda \tilde \rho)}{1-(\lambda -x) \tilde \rho} \, .
\label{current_rod_x}
\end{equation}
The choice of the parameter $x$ implicitly fixes a length scale, determining the curvature of the function $\omega(r)$. The probability for a rod to leave its original position is determined by comparing its headway $r$ to the length $r+x$. Choosing $x=\lambda$ the ratio $\frac{r}{r+\lambda}=\frac{n}{n+l}$ in average measures the local hole density of the lattice system. While a particle on the discrete lattice only hops if it encounters a hole in front due to exclusion, this extra hopping condition is artificially imposed on the hopping rates of the rods. The resulting mass current for $x=\lambda$ is symmetric and formally identical to the ASEP-current:

\begin{equation}
j^{mass}=\lambda j = \tilde \rho^c (1-\tilde \rho^c)\,.
\end{equation}

Choosing $x \ne \lambda$, equation (\ref{generalrates}) yields a non-symmetric current-density relation, especially for $x=1$ the $\ell$-ASEP relation.\\

This means that the length scale fixed by $x$ in the continuous system takes over the role of the lattice spacing in the discrete model. The rods are equally long as this unit in the case $x=\lambda$ for the rod system and the case $\ell=1$ for the $\ell$-ASEP, resulting in a symmetric current. They measure $\lambda$ times and $\ell$ times respectively the length unit for $x=1$ or $\ell \ge 1$. In the case of the rods their real length $\lambda$ is essential, and the current is a function of the volume density. The length parameter $\ell$ of the $\ell$-ASEP particles instead indicates the number of covered sites, therefore the formally equivalent current is a function of the density per lattice site.

\subsection{Lattice simulations of continuum gases}
\label{simulations}
\begin{figure}
\begin{center}
\epsfig{file=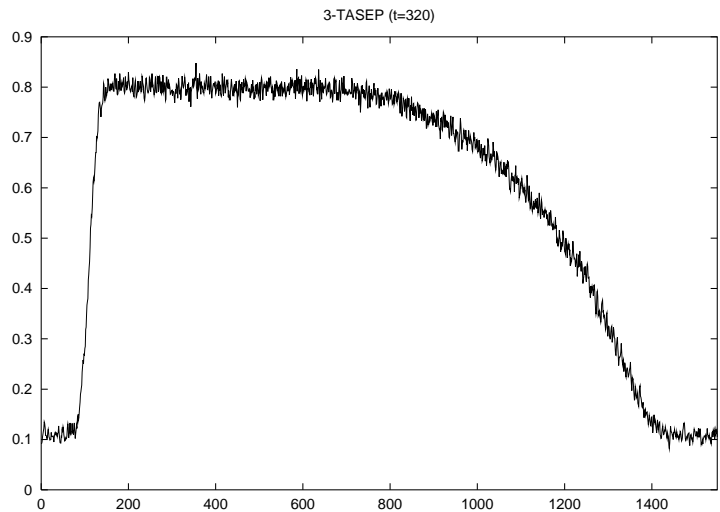, width=7cm, angle=270}
\caption{\label{mc3moving} Evolved density profile after 320 MC steps of an initial step profile of non-maximal density for the 3-TASEP on a periodic lattice of $1500$ sites.}
\end{center}
\end{figure}

\begin{landscape}
\begin{figure}[h!]
\epsfig{file=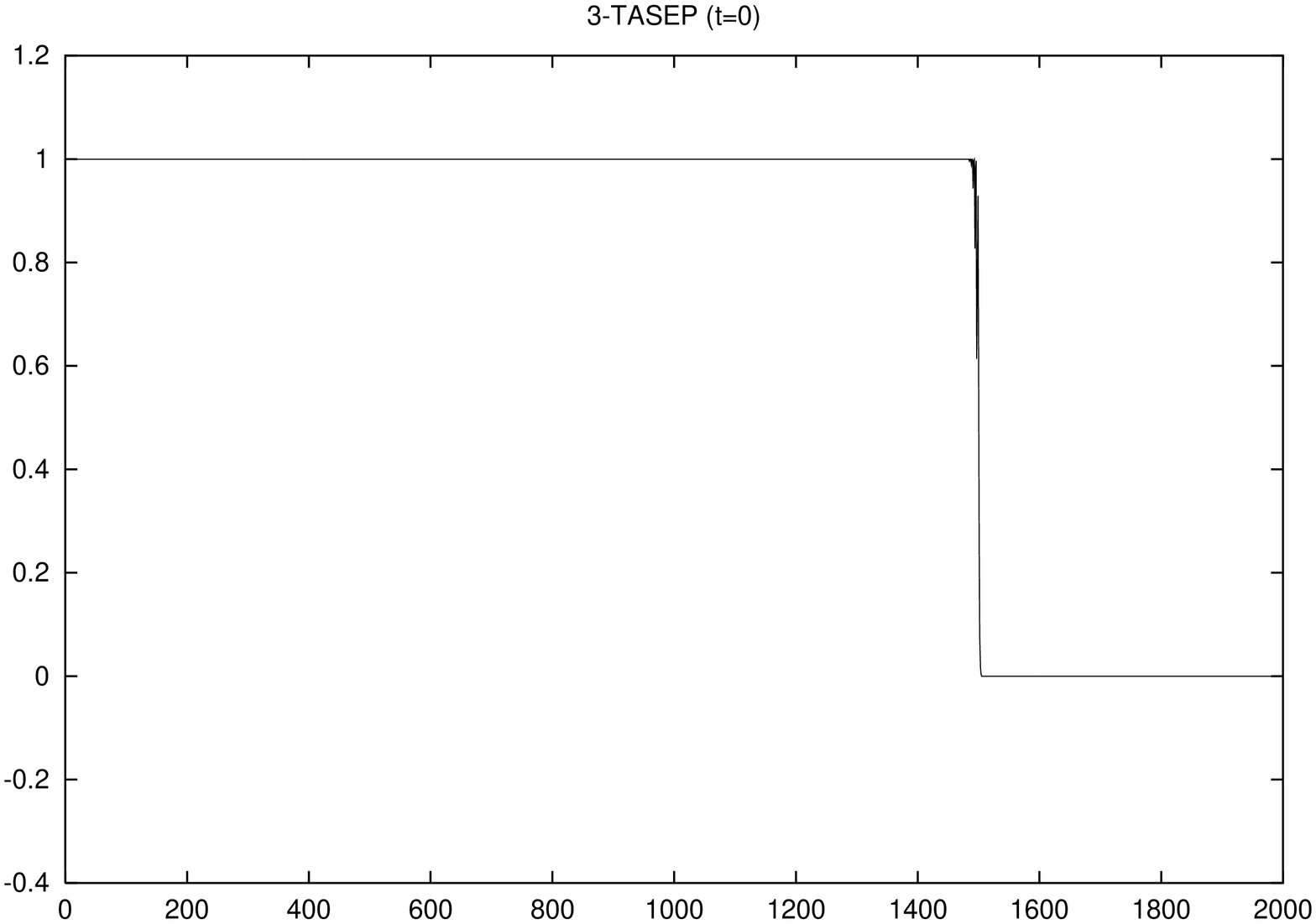, height=10.5cm, angle=270}
\epsfig{file=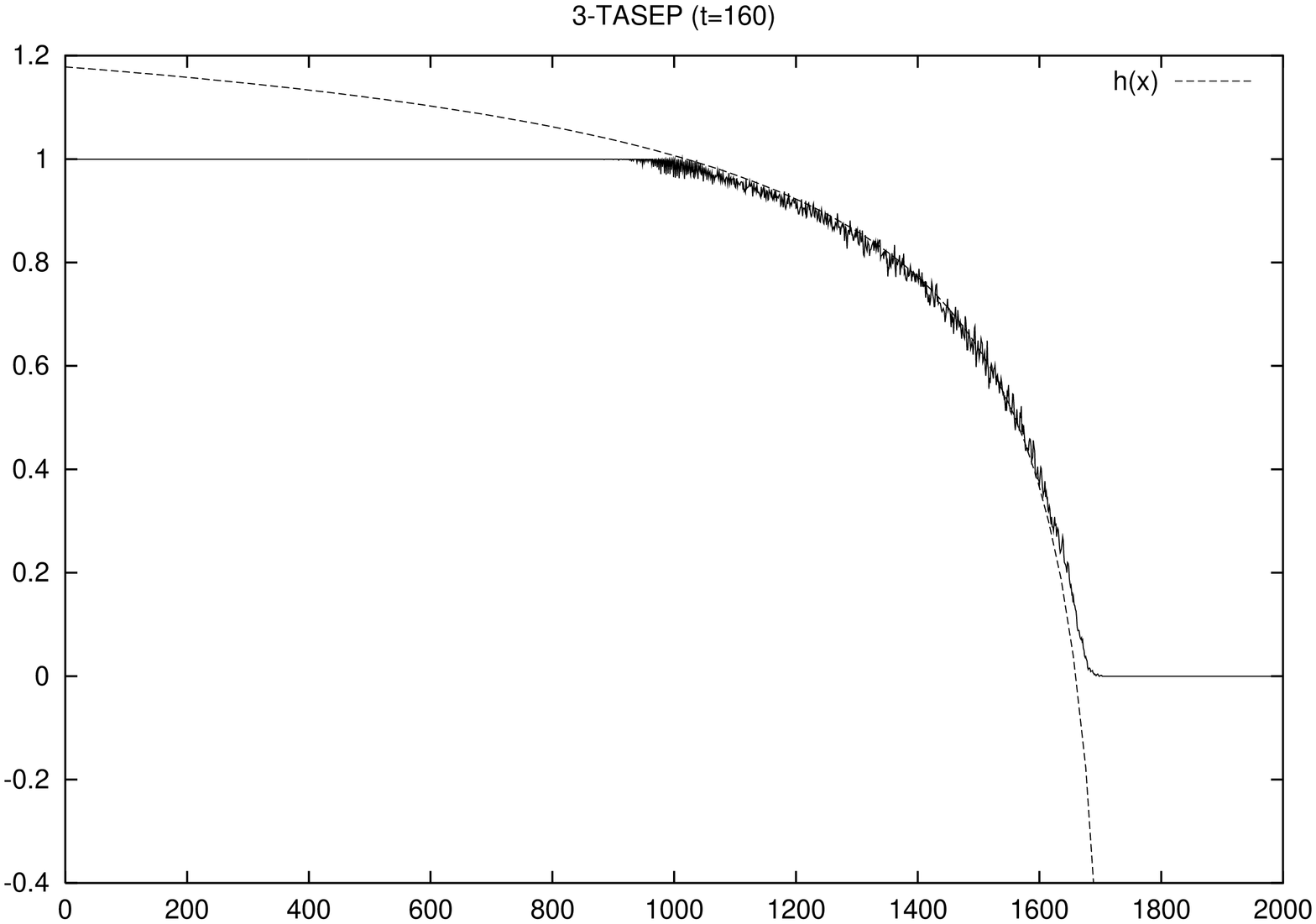, height=10.5cm, angle=270}
\epsfig{file=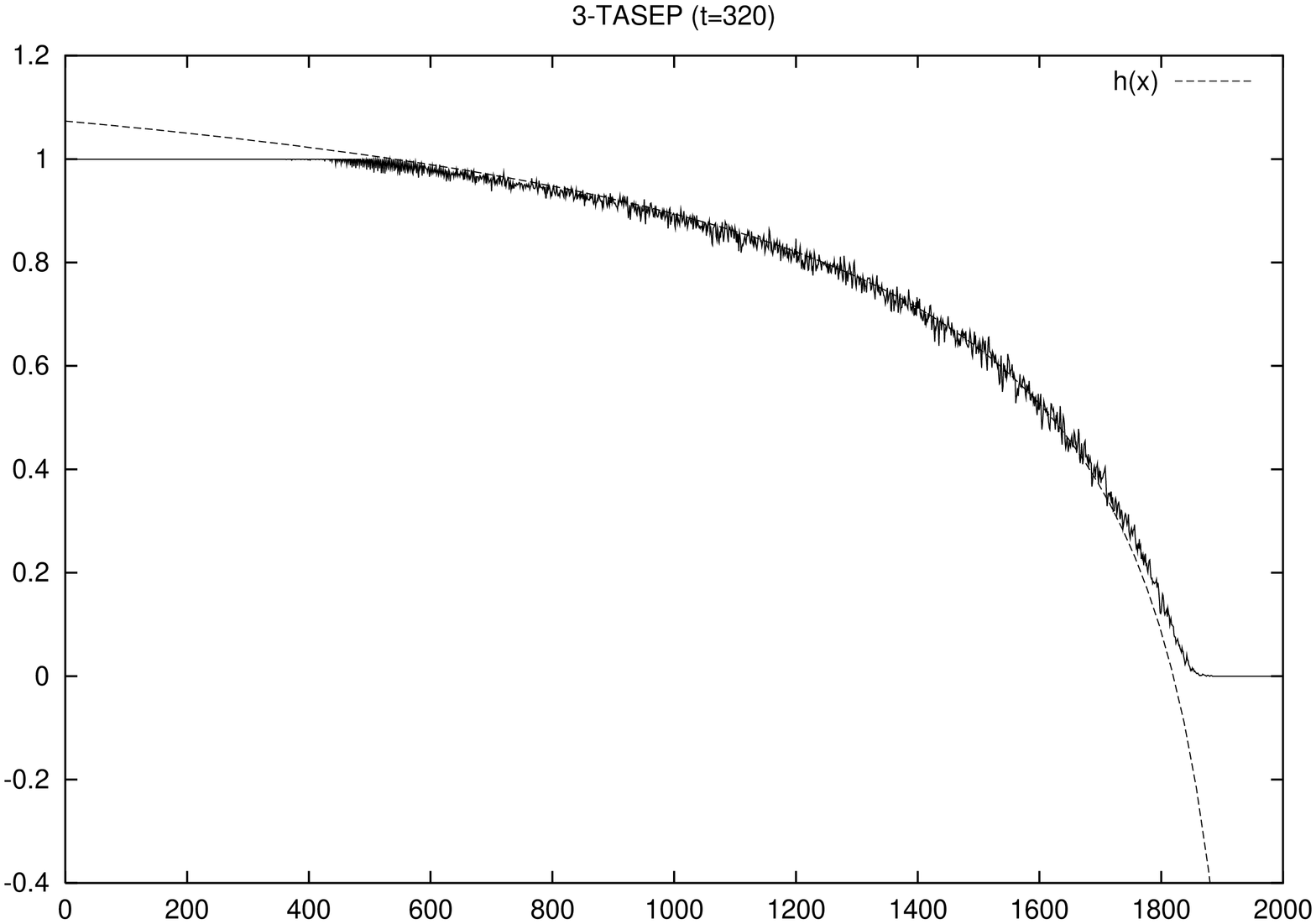, height=10.5cm, angle=270}
\epsfig{file=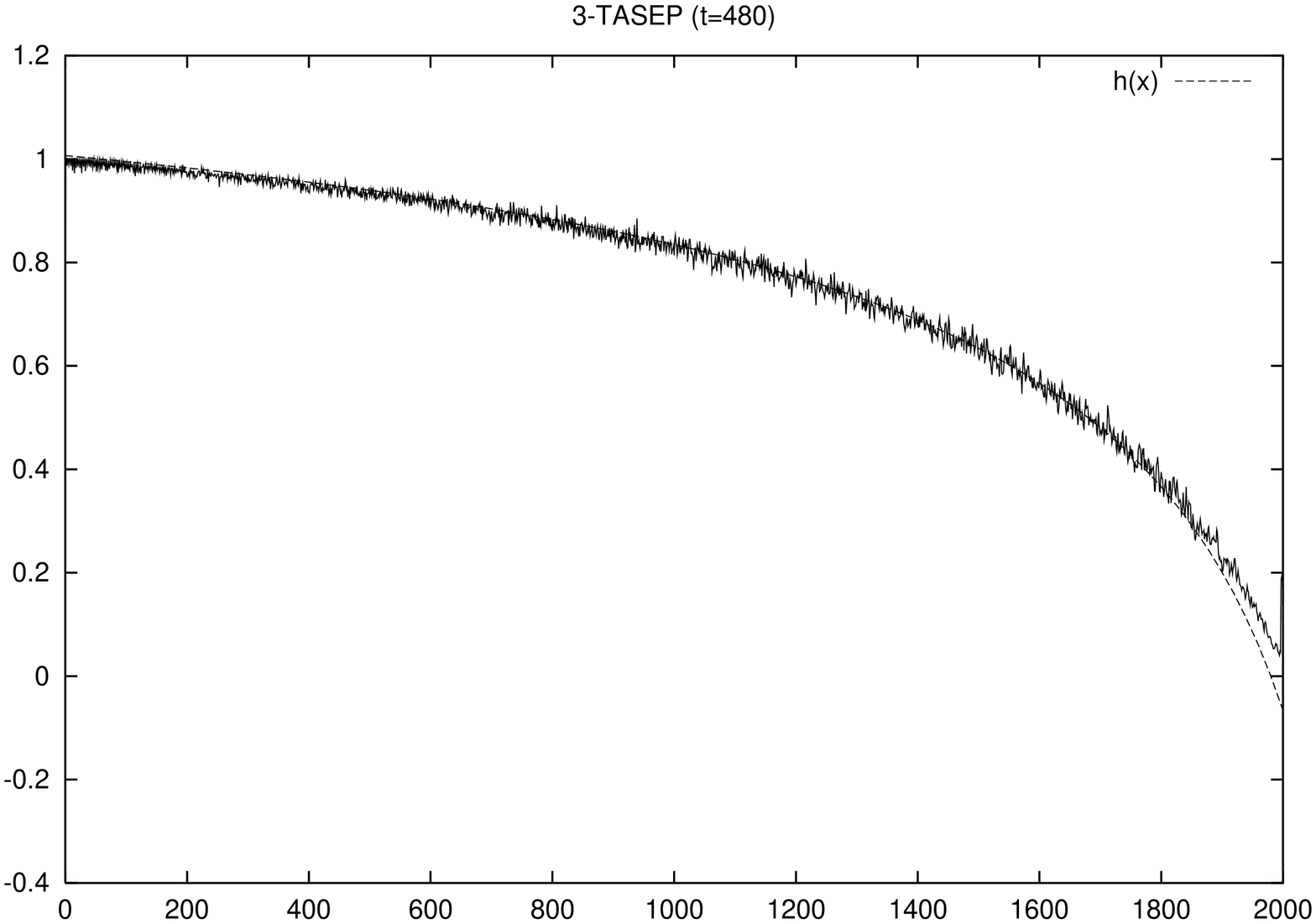, height=10.5cm, angle=270}
\caption{\label{mc3}Density evolution of the ($\ell=3$)-TASEP: The coverage density is plotted on the $y$-axis against the lattice site numbers $0 \cdots 1999$ of a periodic lattice on the $x$-axis. Snapshots from the simulation data are depicted for every $160$ MC steps ($t=0 , 160 , 320 , 480$). The dashed line indicates the theoretical rarefaction wave solution of the density profile. For better graphical illustration it is plotted over a bigger range than the one where it is a valid solution. In its definition range it perfectly matches the simulation data.}
\end{figure}
\end{landscape}

MC simulations are a standard numerical tool for predictions of the evolution of discrete models. In the previous section, the correspondence of a continuous to a discrete system (the $\ell$-ASEP) has been established. This means that one may now employ such numerical approaches as exact simulations of a continuum system. Figures \ref{mc3moving} and \ref{mc3} result from Monte Carlo simulations of the $3$-TASEP as a chosen specification ($\ell=3$, $B=1$, $S=0$) of the $\ell$-ASEP. They illustrate the temporal evolution of a hard rod gas on a periodic lattice with the current-density relation (\ref{current_rod_x})
\begin{equation*}
j(\tilde \rho)=\frac{\tilde \rho (1-3 \tilde \rho)}{1-2 \tilde \rho}
\end{equation*}
starting with initial profiles of the form
\begin{equation}
\tilde \rho^c(x,t=0) = \left\{ \begin{array}{r@{\quad:\quad}l}
1 (0.8) & x < \frac{2}{3} \\
0 & x \ge \frac{2}{3}\,. \end{array} \right.
\label{initialMC}
\end{equation}
A general form of this PDE has been solved analytically in section \ref{TASEPsolutions}. The numerical approach presented here involves MC simulations performed for an ensemble of $1000$ lattices of length $L=2000$ each. Figure \ref{mc3} depicts different stages of the MC evolution of this discrete system where the initial profile (\ref{initialMC}) is realized by a configuration with occupation of sites $0 \ldots 1499$ by $500$ $3$-mers and vacant sites $1500 \ldots 1999$. As periodic boundary conditions are applied, the initial profile at $t=0$ is a \emph{step with two sharp discontinuities} at $0$ and $1499$. For $t >0$, hopping takes place according to $\ell$-TASEP dynamics (a particle hops to the right with rate $1$ provided the lattice site is empty).
One observes that the left discontinuity remains stable until it is reached by the particles coming from the break-up of the right edge of the step. At $t=480$ the equilibration process towards the stationary uniform density profile $\rho \equiv \frac{1}{2}$ has begun. Snapshots were taken at four times in the simulation, every $160$ MC steps starting at $t=0$ after the lattice initialization, in order to show the softening out of the right edge of the initial step. In contrast to the TASEP ($\ell=1$), there is no \emph{linear} decrease in the density profile but it takes the form which was already proposed by the rarefaction solution of the PDE (\ref{pde2}) in section \ref{TASEPsolutions}. For a better comparison of simulation data and theoretical predictions, the rarefaction curve (\ref{rarefactioncurve}) is overplotted as a dashed line for times $t>0$ in figure \ref{mc3}. In its region of validity, linking high- and low-density regime, it matches the MC data very well. The remaining fluctuations are due to the discreteness of the MC system and the small ensemble number chosen for the averaging process. Figure \ref{mc3moving} is a snapshot at $t=320$ from the MC evolution of a similar but non-maximal initial step profile under $3$-TASEP dynamics as a stable shock. The left discontinuity moves as a stable shock with constant velocity towards the right. Except for stochastic fluctuations, the shock is absolutely sharp.\\
\\
The simulations performed match the expected results and confirm theory. The MC simulations, originally a tool to predict the behaviour of the discrete $\ell$-(T)ASEP, also describe a continuum system of long-range interacting hard rods. The existence of stable shock solutions for this hard rod gas is demonstrated in the simulation pictures.

\section{Summary and Conclusions}

The construction of continuum hard rod gases has been achieved by appropriate scaling of and performing limits on discrete lattice models. For the example of the $\ell$-ASEP as an exclusion process (model $A$) and a generalized process with additional long-range interaction (model $B$), two steps were performed: Firstly, in the hard rod limit the length of the extended particles in discrete space was rescaled to be infinite on lattice scale, but to stay finite on a macroscopic scale of continuous space. Secondly the hydrodynamic limit of those hard rods was gained by another coarse-graining procedure. The continuous models achieved in this way do not resemble the hydrodynamic properties of the discrete models from where they arise. More remarkably instead, the hydrodynamic hard rod limit of model $B$ yields exactly the same coarse-grained current-density relation as the hydrodynamic limit of the discrete model $A$.
All basic physical properties of the $\ell$-ASEP (hydrodynamic behaviour and phase transitions \cite{shaw1,lak,schoen,sch3}) are transferable to a continuum system of hard rods with a certain long-range interaction. The methodology behind this correspondence may be exploited for many other types of (lattice) models and inter-particle interactions. Given an empirical current-density relation of some real physical system, the established formalism allows for the construction of a lattice model which exactly reflects the hydrodynamic properties of the continuum system. This opens up new possibilities for the investigation of continuum systems with analytical and in particular numerical tools.\\
\\
{\bf Acknowledgements}\\
\\
I would like to thank Gunter Sch{\"u}tz for proposing this problem and for many
valuable discussions.

\begin{appendix}
\section{Generalized mapping between ZRP and exclusion processes}
Some key steps of the calculation of the HD equation of a generalized $\ell$-ASEP with any type of interaction which can be mapped onto a ZRP as described in section \ref{zrpproperties}, are given. One starts with equation (\ref{HDZRP}).
In order to rewrite its r.h.s in terms of $\rho$, one first carries out the outer derivatives of $z(c(y,t))$, yielding
\begin{equation}
\partial_t c(y,t)=aB\frac{\partial z}{\partial c}\partial_y c(y,t) + a^2S \left[ \frac{\partial^2 z}{\partial c^2}(\partial_y c(y,t))^2 + \frac{\partial z}{\partial c}(\partial_{yy} c(y,t)) \right].
\end{equation}
Secondly one introduces a function $G(\rho)$ which is substituted for $z(c)=z(c(\rho))$, leading to
\begin{eqnarray*}
\partial_t c(y,t)=&- &aB(\rho^2)\partial_{\rho}G(\rho) \partial_y c(y,t) \\
                  &+& a^2S \left[ 2\rho^3\partial_{\rho}G(\rho)+\rho^4\partial_{\rho \rho}G(\rho) \right] (\partial_y c(y,t))^2 \\
                  &-&a^2S\rho^2\partial_{\rho} G(\rho)(\partial_{yy} c(y,t))
\end{eqnarray*}
and thirdly replaces the derivatives of $c(y,t)$ by explicit expressions in $\rho(x,t)$ resulting in
\begin{eqnarray*}
\partial_t c(y,t)=& &aB(\rho^2)\partial_{\rho}G(\rho) \left[ \frac{\rho_x}{\rho^3}+\frac{a}{2}\frac{(\rho_x)^2}{\rho^5} \right] \\
                  &+& a^2S \left[ 2\rho^3\partial_{\rho}G(\rho)+\rho^4\partial_{\rho \rho}G(\rho) \right] \left[ \frac{(\rho_x)^2}{\rho^6} \right] \\
                  &-&a^2S\rho^2\partial_{\rho} G(\rho) \left[ -\frac{\rho_{xx}}{\rho^4}+\frac{3(\rho_x)^2}{\rho^5} \right].
\end{eqnarray*}
Repeating all three steps for the l.h.s.
\begin{equation*}
\partial_t c(y,t)=\frac{\partial c}{\partial \rho} \left[ \frac{\partial \rho}{\partial t}+ \rho_x \frac{\partial x}{\partial t} \right]
\end{equation*}
where $\frac{\partial x}{\partial t}$ is determined by
\begin{equation*}
\frac{\partial x}{\partial t} = aBz(y,t) + a^2S\partial_y z(y,t)- \frac{a^2}{2}B\partial_y z(y,t)
\end{equation*}
one obtains the generalized hydrodynamic equation for particles with exclusion interaction and headway-dependent hopping rates:
\begin{equation*}
\partial_t \rho + aB \left[ G(\rho)+\rho \partial_{\rho} G(\rho) \right] \rho_x + a^2S \left[ \partial_{\rho \rho}G(\rho)(\rho_x)^2 + \partial_{\rho}G(\rho)\rho_{xx} \right]=0.
\end{equation*}
Thus:
\begin{equation}
\partial_t \rho + aB\partial_x[\rho G(\rho)] + a^2S\partial_{xx}G(\rho)=0
\label{vZRPgeneral}
\end{equation}
and
\begin{equation}
j(\rho)=aB\rho G(\rho) + a^2S\partial_{x}G(\rho).
\end{equation}
The above formally introduced function $G(\rho)$ has the physical interpretation of the average particle velocity for a normalized hopping asymmetry $(p-q)=1$, as in lowest order on the Euler scale the following relation holds true:
\begin{equation}
v(\rho)=\frac{\partial j}{\partial \rho}=G(\rho).
\end{equation}

\end{appendix}

\vspace*{1.5 cm}

\begin{landscape}
\begin{tabbing}
\underline{\textbf{ (a) $\ell$-ASEP}} \hspace{10cm}                        \= \underline{\textbf{(b) $\lambda$-rods}} \\
                                                \>                  \\
$L$ lattice sites with $N$ particles, each covering $\ell$ sites.
                   \>  chain of length $\Lambda$ with $N$ rods of length $ \lambda $. \\
densities $\rho = \frac{N}{L}$, $\rho^c = \ell \frac{N}{L}$
                   \> volume densities $\tilde \rho = \frac{N}{\Lambda}$, $\tilde \rho^c = \lambda \frac{N}{\Lambda}$ \\

{\bf (B)} HD limit: $\rho, \rho^c , \ell  \equiv const.$, $a \rightarrow 0, L,N \rightarrow \infty $                                                                                                                                                  \> {\bf (B)} hard rod limit: $\tilde \rho, \tilde \rho^c, \lambda ,\Lambda  \equiv const.$, $a \rightarrow 0, \ell \rightarrow \infty $\\
                                                \>{\bf (C)} HD limit: $\lambda \rightarrow 0$, $\Lambda , N \rightarrow \infty$, $\frac{N}{\Lambda} \equiv const$ \\
\end{tabbing}
\begin{figure}[h!]
\epsfig{file=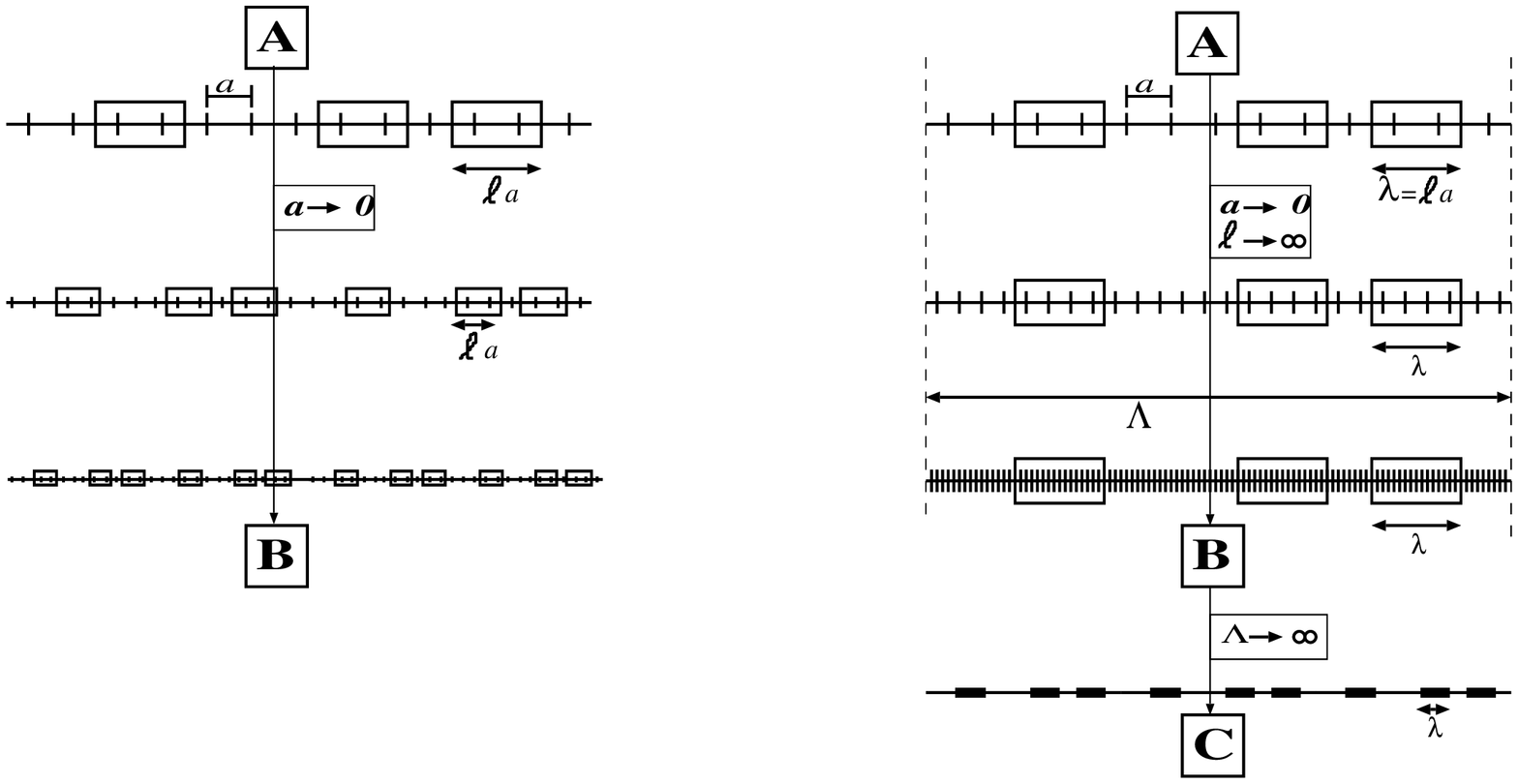, height=20cm, width=8cm,  angle=270}
\caption{\label{continuumlimit}Limiting procedures for $\ell$-ASEP and $\lambda$-rods (A $\rightarrow$ B: continuum limit; B $\rightarrow$ C: HD limit for $\lambda$-rods).}
\label{continuumlimit}
\end{figure}
\end{landscape}

\end{document}